\documentclass[12pt,preprint]{aastex}

\usepackage{natbib}

\slugcomment{Submitted to The Astrophysical Journal}

\begin{document}

\title{Gravitational Radiation from Hydrodynamic Turbulence
 in a Differentially Rotating Neutron Star}


\author{A. Melatos}
\affil{School of Physics, University of Melbourne,
 Parkville, VIC 3010, Australia}
\email{amelatos@unimelb.edu.au}
\and
\author{C. Peralta\altaffilmark{1}}
\affil{Max-Planck-Institut f\"{u}r Gravitationsphysik, 
Albert-Einstein-Institut, Am M\"{u}hlenberg 1, 
D--14476 Golm, Germany}
\altaffiltext{1}{Also at: School of Physics, University of Melbourne,
 Parkville, VIC 3010, Australia}


\begin{abstract}
The mean-square current quadrupole moment associated with vorticity
fluctuations in high-Reynolds-number turbulence in a differentially
rotating neutron star is calculated analytically,
as are the amplitude and decoherence time of the resulting,
stochastic gravitational wave signal.
The calculation resolves the subtle question of whether the signal
is dominated by the smallest or largest turbulent eddies:
for the Kolmogorov-like power spectrum observed in 
superfluid spherical Couette simulations,
the wave strain is controlled by the largest eddies,
and the decoherence time approximately equals 
the maximum eddy turnover time.
For a neutron star 
with spin frequency $\nu_{\rm s}$ and Rossby number ${\rm Ro}$,
at a distance $d$ from Earth,
the root-mean-square wave strain reaches
$h_{\rm RMS} \approx 
 3\times 10^{-24}\, {\rm Ro}^3 (\nu_{\rm s} / 30\,{\rm Hz})^3
 (d/{\rm 1\,kpc})^{-1}$.
Ordinary rotation-powered pulsars 
($\nu_{\rm s} \lesssim 30\,{\rm Hz}$,
 ${\rm Ro} \lesssim 10^{-4}$)
are too dim to be detected by the current generation of 
long-baseline interferometers.
Millisecond pulsars are brighter;
for example,
an object born recently in a Galactic supernova or 
accreting near the Eddington rate can have
$\nu_{\rm s} \sim 1\,{\rm kHz}$, ${\rm Ro}\gtrsim 0.2$,
and hence
$h_{\rm RMS} \sim 10^{-21}$.
A cross-correlation search can detect such a source in principle,
because the signal decoheres over the time-scale 
$\tau_{\rm c} \approx 1 \times 10^{-3} \, {\rm Ro}^{-1}
 (\nu_{\rm s} / 30\,{\rm Hz})^{-1} \, {\rm s}$,
which is adequately sampled by existing long-baseline interferometers.
Hence hydrodynamic turbulence imposes a fundamental
noise floor on gravitational wave observations of neutron stars,
although its polluting effect may be muted by partial decoherence
in the hectohertz band,
where current continuous-wave searches are concentrated,
for the highest frequency (and hence most powerful) sources.
This outcome is contingent on the exact shape 
of the turbulent power spectrum,
which is modified by buoyancy and anisotropic global structures,
like stratified boundary layers, 
in a way that is understood incompletely even in laboratory situations.
\end{abstract}



\newpage 


\section{Introduction
 \label{sec:tur1}}
Shortly after the discovery of radio pulsars, speculation arose
that the superfluid interior of a differentially rotating
neutron star is turbulent
\citep{gre70}.
Since then, 
the theme has resurfaced intermittently during
the quest to understand pulsar rotational irregularities,
like glitches and timing noise
\citep{and78,tsa80}.
Neutron star turbulence can be hydrodynamic,
taking the form of a Kolmogorov-like cascade of {\em macroscopic} eddies
at high Reynolds numbers
\citep{per05,per06a,mel07,per08}.
Complicated vorticity patterns of this sort are observed
in terrestrial experiments on spherical Couette flow,
which undergo transitions to nonaxisymmetric flow states
at high Reynolds numbers,
e.g.\ spiral, shear herringbone, or Taylor-G\"{o}rtler vortices
\citep{bel79,buh90,jun00,sha01,nak02a,nak02b,nak05a,per06b,per08},
relaminarization
\citep{nak05b},
or Stewartson layer disruption
\citep{hol03,hol04,hol06,wei08}.
Neutron star turbulence can also be quantum mechanical,
comprising a self-sustaining tangle of quantized {\em microscopic} vortices 
\citep{gla74,jou04,jou06},
excited by bulk two-stream instabilities
\citep{and07},
interfacial two-stream instabilities
\citep{bla02,mas05},
or meridional circulation
\citep{per05,per06a,mel07}.
In general, macroscopic and microscopic superfluid turbulence
appear to trigger each other; 
it is an unsolved, chicken-or-egg question
as to which comes first
\citep{bar01,tsu09}.

Turbulence powered by differential rotation is axisymmetric
when averaged over long times but nonaxisymmetric instantaneously.
Turbulent flows therefore emit stochastic gravitational waves. 
In an incompressible fluid, the waves arise
mainly from current quadrupole (and higher multipole) source terms.
In a compressible fluid, the mass multipoles also contribute;
indeed, they can dominate,
e.g.\ during post-glitch Ekman pumping in a neutron star
\citep{van08}.
Recently,
\citet{per06b} pointed out that there exists a fundamental 
theoretical uncertainty
regarding the shape and strength of the gravitational wave signal
emitted by hydrodynamic turbulence.
The mechanical stress-energy in Kolmogorov-like turbulence 
is contained mostly in large eddies near the stirring scale.
Naively, therefore, one might expect the gravitational wave signal
to look like a `dirty sinusoid',
which reflects circulation on the largest scales and
decoheres in approximately one rotation period.
However, the instantaneous wave strain is proportional to
the second time-derivative of the stress-energy tensor,
and this quantity is greatest for small eddies
near the dissipation scale,
which turn over most quickly.
If the latter effect dominates, one might expect the signal
to resemble white noise.
Of course, large eddies match better to low-order multipoles
than small eddies, and low-order multipoles typically dominate
the gravitational wave strain far from the source
\citep{was09}.
A careful calculation is therefore required to select between
the various possibilities and reliably estimate the detectability
of the signal.

In this paper, we undertake such a calculation by combining
the formalism of \citet{kos02} and \citet{gog07},
developed to calculate the gravitational radiation from a
turbulent, first-order phase transition in the early Universe,
together with the formalism of \citet{was09},
developed to calculate the gravitational radiation from
nonaxisymmetric vorticity fluctuations in neutron stars,
e.g.\ due to clusters of quantized superfluid vortices.
In \S\ref{sec:tur2},
we analyze global hydrodynamic simulations of incompressible,
shear-driven neutron star turbulence to extract the 
vorticity correlation function,
which feeds into the statistics of stress-energy fluctuations
in the source.
We then calculate analytically the current multipole moments,
root-mean-square (RMS) wave strain,
and decoherence time
of the resulting, stochastic gravitational wave signal in \S\ref{sec:tur3}.
The results are applied in \S\ref{sec:tur4} to estimate
the detectability of the signal,
e.g.\ with long-baseline interferometers
like the Laser Interferometer Gravitational-Wave Observatory (LIGO),
and its polluting effect on continuous-wave searches currently
under consideration.
Hydrodynamic turbulence imposes a fundamental, quantifiable
noise floor on gravitational wave observations of neutron stars.
Astrophysical implications, including the rate of gravitational wave
braking in different types of neutron stars,
are briefly canvassed.

\section{Turbulent vorticity correlations
 \label{sec:tur2}}
Let ${\bf \omega}({\bf x},t)$ be a turbulent vorticity field
which fluctuates stochastically with position ${\bf x}$
and retarded time $t$ in the source.
The gravitational wave strain generated by
the $(l,m)$-th current multipole
is proportional to the $l$-th time derivative of ${\bf \omega}({\bf x},t)$
integrated over the source volume,
as discussed by \citet{tho80} and in \S\ref{sec:tur3b}.
If the turbulence is stationary,
the wave strain averages to zero over times longer than
the maximum eddy turnover time.
However, the RMS wave strain is not zero.
It is proportional to the autocorrelation function
$\left\langle 
  \omega_i({\bf x},t) \omega_j^\ast({\bf x}',t')
 \right\rangle$
integrated over the source volume.
Here and elsewhere,
angular brackets denote the usual ensemble average.
Expressing ${\bf \omega}({\bf x},t)$ in terms of its
spatial Fourier transform, we can write
\begin{equation}
 \left\langle 
  \omega_i({\bf x},t) \omega_j^\ast({\bf x}',t')
 \right\rangle
 =
 \epsilon_{ilm} \epsilon_{jpq}
 \int \frac{d^3{\bf k}}{(2\pi)^3}
 \int \frac{d^3{\bf k}'}{(2\pi)^3}
 e^{i{\bf k}\cdot{\bf x}-i{\bf k}'\cdot{\bf x}'}
 k_l k_p'
 \left\langle
  v_m({\bf k},t) v_q^\ast({\bf k}',t')
 \right\rangle~,
 \label{eq:tur3}
\end{equation}
where ${\bf v}({\bf x},t)$ is the turbulent velocity field
satisfying 
${\bf\omega}({\bf x},t) = {\rm curl}\,{\bf v}({\bf x},t)$.
For stationary turbulence, 
$\left\langle 
  \omega_i({\bf x},t) \omega_j^\ast({\bf x}',t')
 \right\rangle$
depends on $t$ and $t'$ only through the combination
$\tau=t'-t$.

\begin{figure}
\plotone{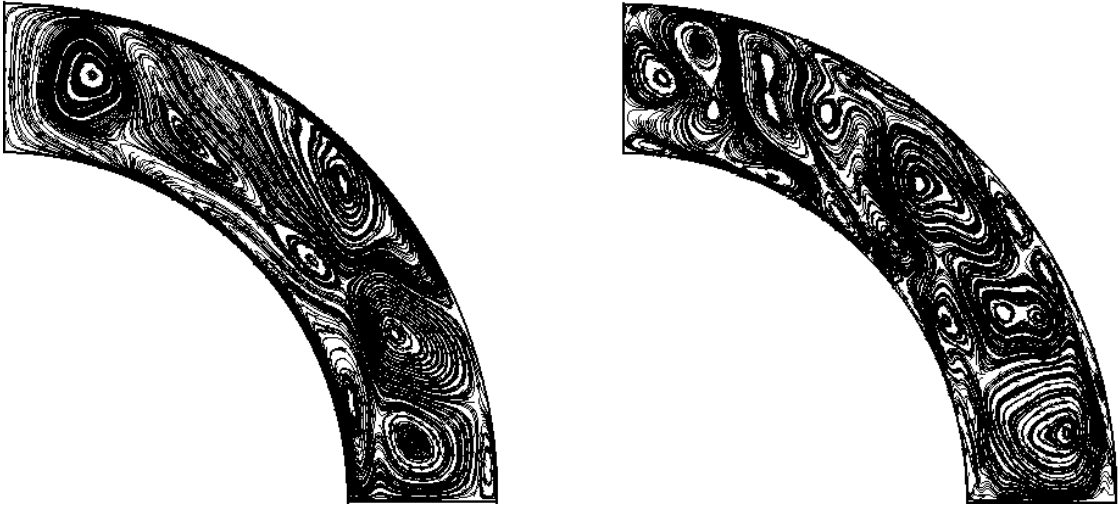}
\caption{
Hydrodynamic turbulence in a two-component, incompressible,
HVBK superfluid in a differentially rotating shell,
with dimensionless thickness $\delta=0.3$,
Rossby number ${\rm Ro}=\Delta\Omega/\Omega=0.1$,
and Reynolds number ${\rm Re}=3\times 10^4$.
Parameters are defined in \citet{per08}.
The figure shows a snapshot of the meridional streamlines
of the two components at time $t=48\Omega^{-1}$,
taken from the numerical simulations in \citet{per08}.
The rotation axis points vertically.
{\em Left panel.} Viscous component.
{\em Right panel.} Inviscid component.
}
\label{fig:tur1}
\end{figure}

Global, three-dimensional, numerical simulations of neutron star turbulence
driven by differential rotation indicate that the turbulence
is approximately isotropic and stationary
once the Reynolds number ${\rm Re}$ exceeds $\sim 10^4$
\citep{per05,per06a,mel07,per08}.
Figure \ref{fig:tur1} presents meridional streamlines of
the viscous ({\em left panel}) and inviscid ({\em right panel})
components of a two-component, incompressible,
\footnote{The incompressible approximation is acceptable
when the turbulent motions are subsonic,
as in a neutron star
\citep{per05}.}
Hall-Vinen-Bekarevich-Khalatnikov (HVBK) superfluid
\citep{hil77}
in a differentially rotating shell,
with dimensionless thickness $\delta=0.3$,
Rossby number ${\rm Ro}=0.1$,
and Reynolds number ${\rm Re}=3\times 10^4$,
showing a snapshot of the flow at time 
$t=4.8{\rm Ro}^{-1}\Omega^{-1}$,
where $\Omega$ denotes the angular velocity of the stellar surface.
The simulation parameters are defined precisely in \citet{per08},
where the numerical algorithm (pseudospectral collocation) 
is also laid out in detail.
\footnote{
Such simulations typically adopt
no-slip and no-penetration boundary conditions
for the viscous component, perfect slip or no slip for the inviscid component, 
a Stokes flow initially,
and a mutual friction force of the Gorter-Mellink form
appropriate for a quantized vortex tangle
\citep{gor49,gla74,per05,and07}.
}
Although the Reynolds number in Figure \ref{fig:tur1} 
is $\sim 10^7$ times less than in a realistic neutron star
and only just above the threshold for turbulence,
it is already possible to see
that departures from isotropy are limited to the
largest scales, i.e. $\sim R_\ast\delta$,
where $R_\ast$ is the stellar radius.
This is true even when the shear is stronger;
Rossby numbers as high as 0.3 have been investigated.
Additional pictorial examples appear in \citet{per08}.
Isotropy is expected to increase with ${\rm Re}$,
as in other turbulent systems,
but simulations with ${\rm Re} \geq 10^5$,
which would test this claim,
are not feasible computationally at present.

\begin{figure}
\plottwo{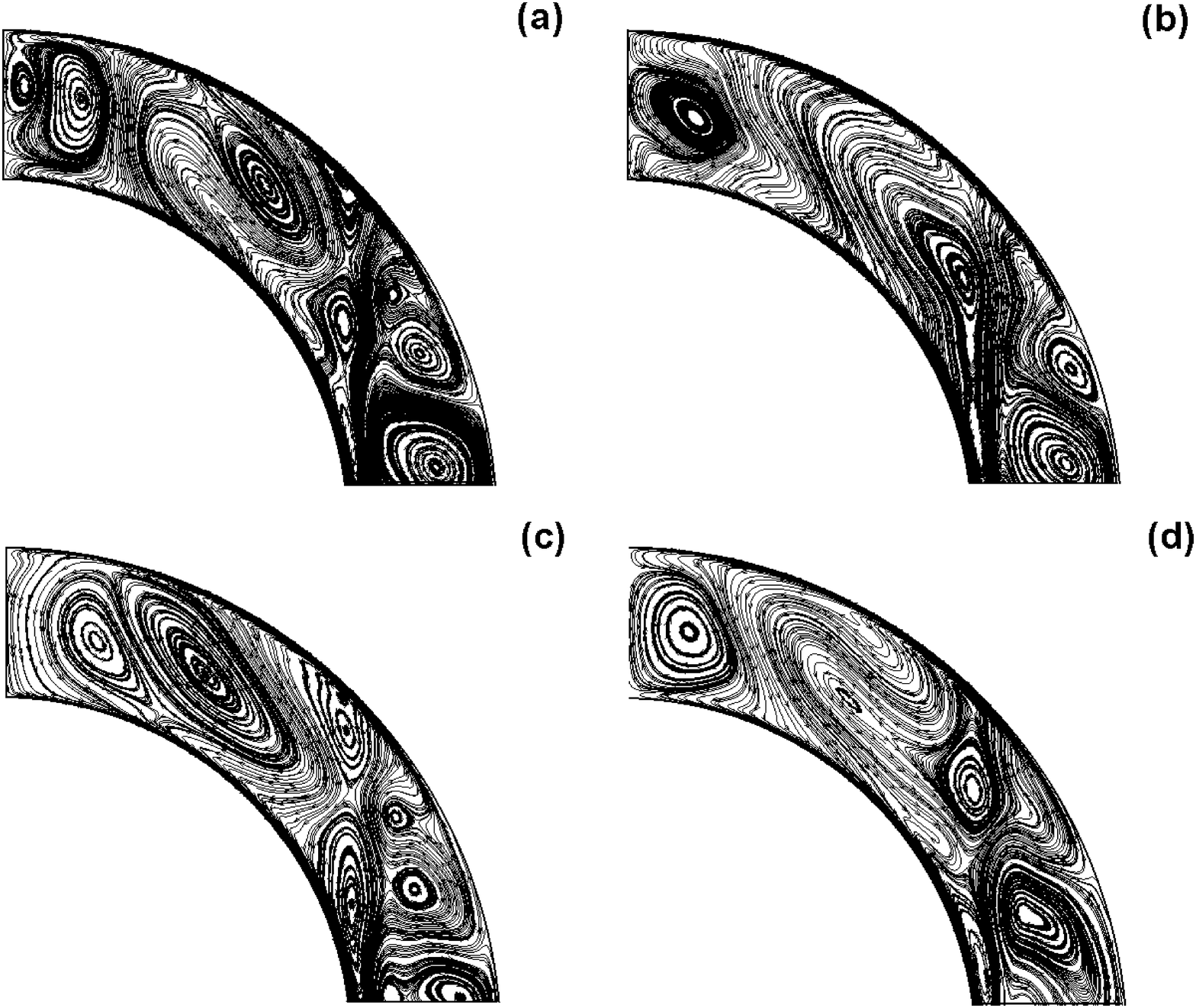}{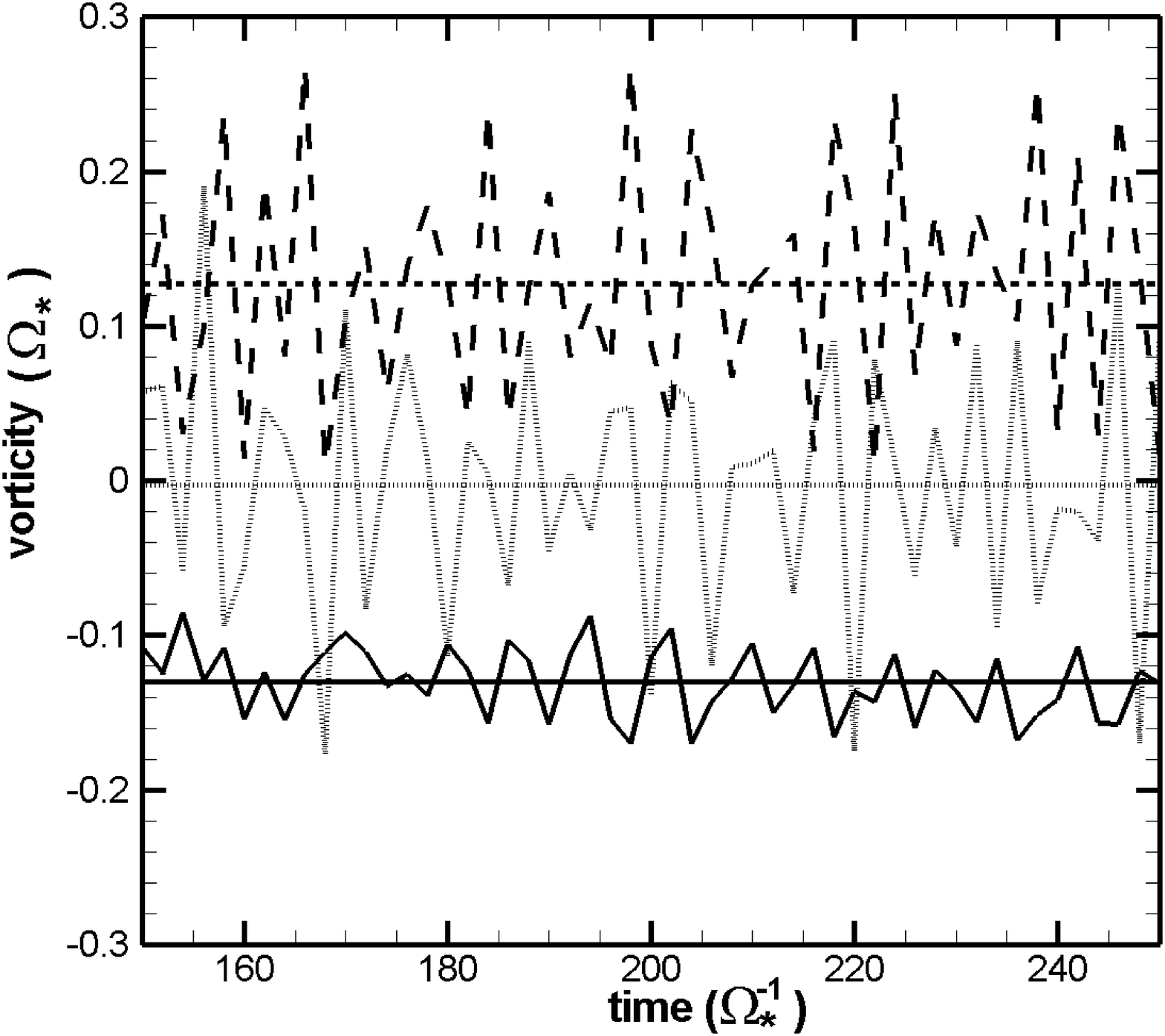}
\caption{
Stationarity of the turbulence in Figure \ref{fig:tur1},
for the same simulation parameters.
{\em Left panel.}
Snapshots of the meridional streamlines of the 
viscous HVBK component at 
($a$) $t=50\Omega^{-1}$, 
($b$) $t=52\Omega^{-1}$, 
($c$) $t=54\Omega^{-1}$, 
and
($d$) $t=56\Omega^{-1}$. 
The streamlines reorganize stochastically
on the time-scale $\Omega^{-1}$.
{\em Right panel.}
Vorticity at a fixed point
($r=0.85R_\ast$, $\theta=\pi/4$, $\phi=0$)
as a function of time 
($150\leq \Omega t \leq 250$).
The spherical polar vorticity components are plotted
after subtracting the rigid body rotation,
viz.\
$\omega_r-2\Omega_r$ (solid curve),
$\omega_\theta-2\Omega_\theta$ (dashed curve),
and
$\omega_\phi$ (dotted curve).
The horizontal lines represent the mean value of
each component over the plotted range.
The vector sum of the means equals
$2(\Delta\Omega){\bf \hat z}$,
the residual differential vorticity.
}
\label{fig:tur2}
\end{figure}

The turbulence in Figure \ref{fig:tur1} is stationary
to a good approximation.
The streamline pattern reorganizes
stochastically on the time-scale $\Omega^{-1}$,
and the velocity components at a fixed point
alternate in sign,
in such a way that the vorticity averages
to the rigid body value $2{\bf \Omega}$ over the long term.
This behavior is summarized in Figure \ref{fig:tur2}.
The left panel shows the meridional streamlines of the viscous
HVBK component at four instants in time,
each separated by $2\Omega^{-1}$.
The eddies in the flow change noticeably in shape, size, and position
throughout the simulation
(and in the inviscid component, which is not plotted).
In the right panel,
we graph all three vector components of the vorticity 
at a mid-latitude point versus time,
after subtracting the rigid body term $2{\bf \Omega}$.
The mean of each component is plotted as a horizontal line
for comparison.
After initial transients die away,
i.e.\ for $t \gtrsim 20\Omega^{-1}$,
the turbulent vorticity fluctuates stochastically and
without bias about its mean value and is independent of $t$.
It has zero mean after adjusting for the
residual differential vorticity $2(\Delta\Omega){\bf \hat z}$.

Isotropic, stationary turbulence has a velocity correlation function
$\left\langle
  v_m({\bf k}) v_q^\ast({\bf k}')
 \right\rangle
=
V (2\pi)^3 {\hat P}_{mq}(k) P(k) 
 \delta({\bf k} - {\bf k}')$,
where $V$ is the total volume of the system (and drops out
at the end of the calculation of any physical observable),
${\hat P}_{mq}(k)=\delta_{mq}-{\hat k}_m{\hat k}_q$
is a projection operator perpendicular to
the wave vector ${\bf k}$,
and $P(k)$ is the power spectrum of the turbulence,
usually a power law.
However, for gravitational wave problems, where we are
interested in temporal fluctuations of the mean-square 
wave strain (and hence mean-square multipole moments),
we are obliged to work with unequal time ($t\neq t'$)
correlators of the kind appearing in (\ref{eq:tur3}).
As a working hypothesis, in this paper, we assume the
standard Kraichnan form for high-${\rm Re}$ turbulence
in three dimensions
\citep{kra59,kos02},
viz.\
\begin{equation}
 \left\langle
  v_m({\bf k},t) v_q^\ast({\bf k}',t')
 \right\rangle
 =
 V (2\pi)^3 {\hat P}_{mq}(k) F(k,t-t') 
 \delta({\bf k} - {\bf k}')~,
\label{eq:tur4}
\end{equation}
with
\begin{equation}
 F(k,\tau) = P(k) \exp[-\pi \eta(k)^2 \tau^2/4]
\label{eq:tur5}
\end{equation}
and
\begin{equation}
 \eta(k)
 =
 (2\pi)^{-1/2} \varepsilon^{1/3} k^{2/3}~.
\label{eq:tur6}
\end{equation} 
In (\ref{eq:tur5}) and (\ref{eq:tur6}),
$\varepsilon$ is the energy dissipation rate per unit enthalpy
(units: ${\rm m^2\,s^{-3}}$),
and $\eta(k)^{-1}$ is the eddy turnover time at wavenumber $k$;
that is, turbulent motions with wavelength $2\pi/k$ in any
given local region decohere over time intervals 
longer than $\eta(k)^{-1}$.

What is $P(k)$ for neutron star turbulence?
Alas, there is little one can say with confidence
about this question, given the impossibility of direct
measurements and the worrying experience with terrestrial systems, 
where idiosyncratic features are often imprinted on $P(k)$ 
by boundary layers and other unavoidable global structures 
even in simple systems
(see next paragraph).
Nevertheless, in order to make progress,
we assume that $P(k)$ is a power law,
$P(k) \propto k^{\alpha}$,
and that the power-law exponent is close to the
Kolmogorov value for isotropic, high-${\rm Re}$ turbulence,
$\alpha=-11/3$
\citep{kra59,gog07}.
Following the standard normalization recipe, 
we can then write
\begin{equation}
 P(k) = V^{-1} \pi^2 \varepsilon^{2/3} k^{-11/3}~.
\label{eq:tur7}
\end{equation}
The power law stretches across an inertial range
extending from the wavenumber corresponding to the
stirring scale,
\begin{equation}
 k_{\rm s} = 2\pi/R_\ast,
\label{eq:tur7a}
\end{equation}
up to the wavenumber corresponding to the 
viscous dissipation scale,
\begin{equation}
 k_{\rm d} = [8 \varepsilon/(27\nu^3)]^{1/4},
\label{eq:tur7b}
\end{equation}
where $\nu$ denotes the kinematic viscosity.

\begin{figure}
\plottwo{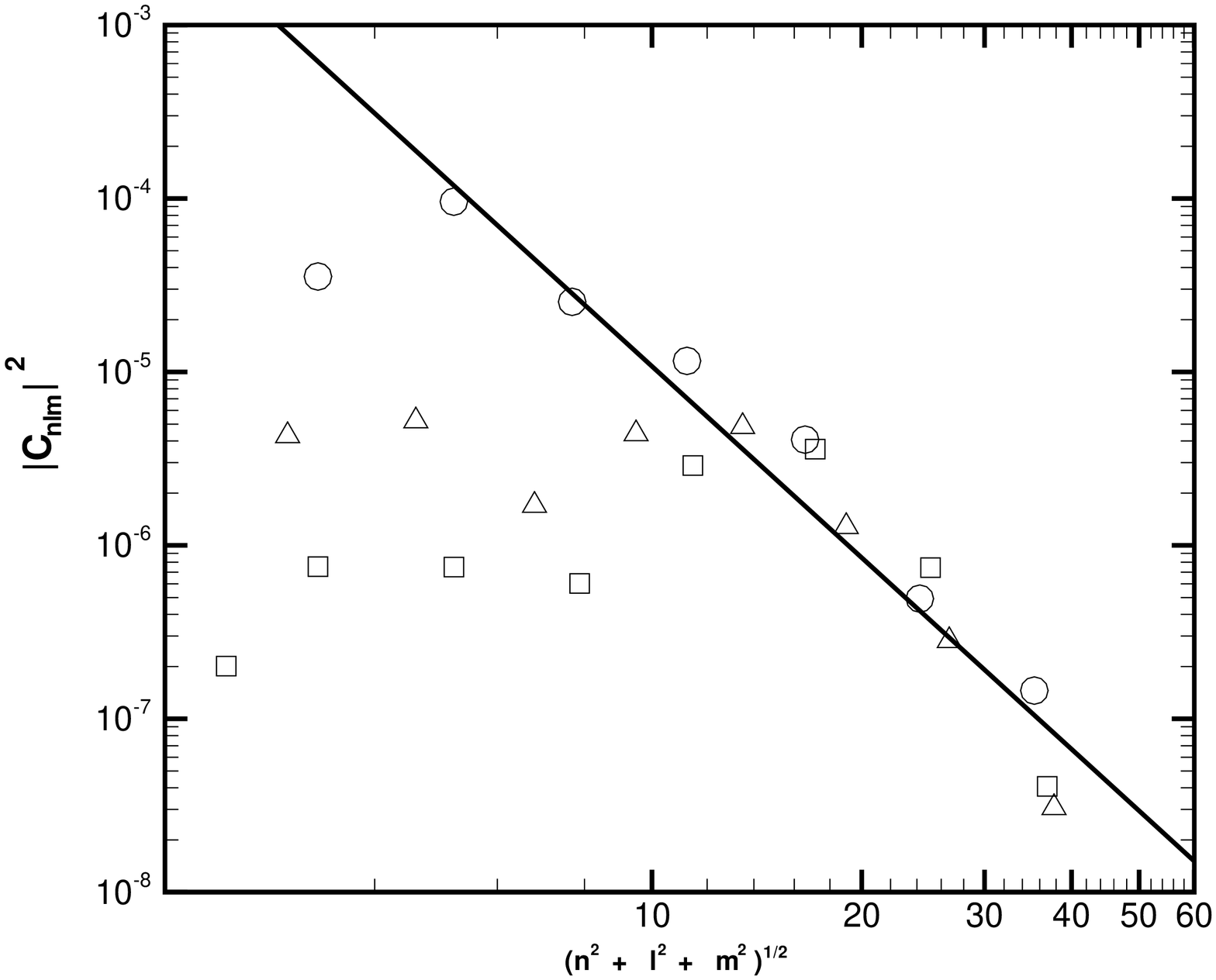}{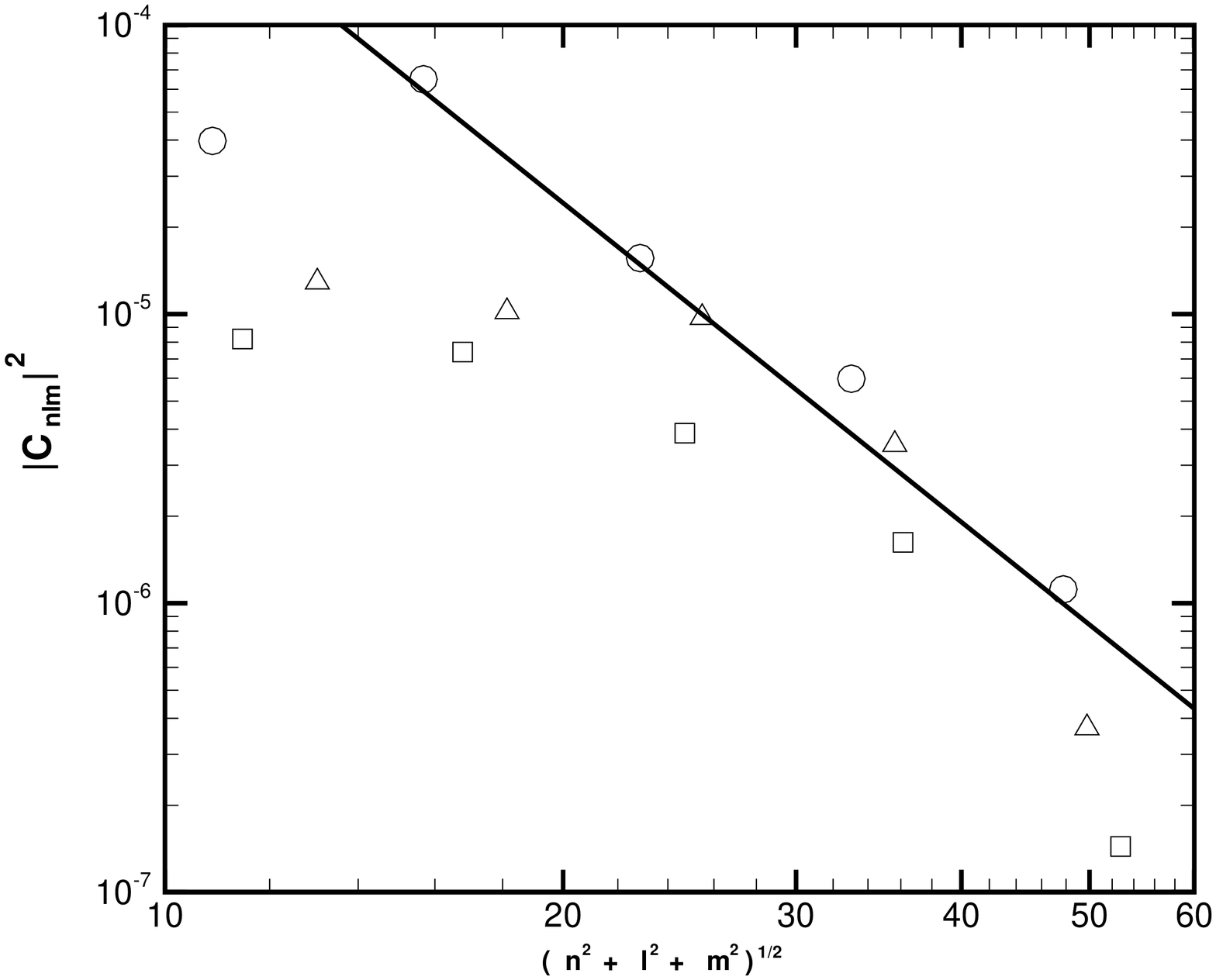}
\caption{
Turbulent power spectrum $P(k)$ (in arbitrary units)
for the simulation parameters in Figure \ref{fig:tur1},
viz.\
$\delta=0.3$, ${\rm Ro}=0.1$, and ${\rm Re}=3\times 10^4$.
The power spectrum is proportional to
$\sum_{n,l,m} |C_{nlm}|^2$,
where $(n,l,m)$ are mode indices in the pseudospectral expansion;
for a given wavenumber bin $[k,k+dk]$,
the sum runs over all indices satisfying
$k\leq (n^2+l^2+m^2)^{1/2} \leq k+dk$
(among the top $10^3$ modes).
A snapshot of the spectrum at $t=50\Omega^{-1}$ is plotted
as a function of normalized wavenumber $kR_\ast/2\pi$,
with logarithmic binning.
It contains three contributions,
$|v_r({\bf k},t)|^2$ ({\em open squares}),
$|v_\theta({\bf k},t)|^2$ ({\em open triangles}),
and
$|v_\phi({\bf k},t)|^2$ ({\em open circles}),
whose sum is proportional to $P(k)$.
The Kolmogorov scaling ({\em solid line}),
with arbitrary normalization, is overplotted for comparison.
Slightly different bins are used for the three terms.
The steady-state differential rotation,
contained in $C_{010}$,
lies off the scale.
{\em Left panel.} Viscous HVBK component.
{\em Right panel.} Inviscid HVBK component.
}
\label{fig:tur3}
\end{figure}

Direct numerical simulations provide reasonable support
for the Kolmogorov scaling
\citep{per08}.
In Figure \ref{fig:tur3}, we construct $P(k)$ for the
velocity field in Figure \ref{fig:tur1}
from the simulation data
by summing the squared pseudospectral coefficients $|C_{nlm}|^2$ 
corresponding to each value of $k$
for the top $10^3$ modes
\citep{per05,per08}.
That is, we plot $P(k)$ as a function of
$kR_\ast/2\pi = (n^2 + l^2 + m^2)^{1/2}$,
where $n$, $l$, and $m$ denote 
radial, latitudinal, and toroidal mode indices respectively.
The viscous and inviscid HVBK components are analyzed in
the left and right panels respectively.
We first note that the toroidal contribution
$|v_\phi({\bf k},t)|^2$ ({\em open circles}) 
dominates $P(k)$, especially at small $k$,
and adheres closely to the Kolmogorov scaling
({\em solid line}).
The poloidal contributions 
$|v_r({\bf k},t)|^2$ ({\em open squares})
and
$|v_\theta({\bf k},t)|^2$ ({\em open triangles})
deviate from the Kolmogorov scaling at small $k$
because isotropy breaks down for the largest eddies,
as noted before;
but, in any case, $P(k)$ is dominated by 
$|v_\phi({\bf k},t)|^2$ at small $k$.
A least-squares fit to $P(k)$ over the inertial range 
in Figure \ref{fig:tur3}
yields $\alpha=-3.52\pm 0.35$ for the viscous HVBK component
($8 \lesssim kR_\ast/2\pi \lesssim 38$)
and $\alpha=-3.55\pm 0.25$ for the inviscid HVBK component
($16 \lesssim kR_\ast/2\pi \lesssim 53$).
These results agree surprisingly well with the Kolmogorov value
$\alpha=-11/3$,
even though the turbulence in Figure \ref{fig:tur1} is not
fully developed,
the inertial range stretches over less than one decade
in the simulations, 
and there are departures from isotropy at small $k$.
We have verified that spectral filtering, 
which is implemented in the numerical solver to enhance its stability
\citep{per05,per08}, 
does not warp $P(k)$ significantly 
for ${\rm Re}=3\times 10^4$.

It is important to reiterate that the scaling (\ref{eq:tur7}) 
applies to isotropic turbulence in the bulk, 
e.g.\ grid turbulence far from any walls.
It is known from many laboratory experiments,
e.g. in wind tunnels, 
that $P(k)$ is modified by the presence of
anisotropic global structures like boundary layers,
\footnote{
Boundary layers in a neutron star are thin, 
e.g.\ ${\rm Re}^{-1/2}$ (surface Ekman layer),
${\rm Re}^{-1/3}$ (Stewartson layer tangent to the core),
or ${\rm Re}^{-2/5}$ (equatorial Ekman layer)
in units of $R_\ast$
\citep{per09}. 
Nevertheless, they influence a large volume of fluid by partitioning
the flow globally into cells,
thereby shaping $P(k)$ at low $k$.
Laboratory experiments also reveal transient,
ribbon-like streamers, which resemble boundary layers,
throughout the body of otherwise isotropic Kolmogorov turbulence.
The streamers generate anomalous Reynolds stresses and
significant instantaneous departures from isotopy at low $k$
\citep{mar01,gan03}.
}
to the point where it may not even be a power law
\citep{sad94}.
The modifications are not merely of academic interest;
we show in \S\ref{sec:tur3} that the amplitude of the
gravitational wave signal from turbulence is sensitive
to the form of $P(k)$,
e.g.\ through $\alpha$.
Unfortunately, calculating $P(k)$ accurately is a
formidable undertaking even in terrestrial situations,
where conditions can be controlled,
let alone in a neutron star.
A voluminous literature exists on turbulent cascades
in shear and viscous boundary layers; 
see, for example, the review by \citet{rob91}.
Experiments that use
stereoscopic particle image velocimetry to measure
the instantaneous velocity field and hence $P(k)$
find that the boundary layer is populated by
coherent structures, 
like hairpin vortices
\citep{gan03}
or wall-wake flows
\citep{per95,mar01},
and anomalous Reynolds stresses
\citep{wie94,gan03},
which are inconsistent with the Kolmogorov model
\citep{sad94}.
The role of stratification
\citep{fer91},
important in a neutron star,
and the interplay between shear and buoyant convection
\citep{moe94},
are also under active investigation.
Resolving these matters lies far outside the scope of
this paper, but it is important to recognize them
and work towards a better understanding over time,
e.g.\ by improving upon the pioneering superfluid
spherical Couette simulations of \citet{per08}.

Buoyancy suppresses radial motion in a neutron star, 
arguably reducing the turbulence to two dimensions.
\citet{kra67} postulated that two-dimensional turbulence develops two inertial ranges: 
a $-5/3$ cascade ($\alpha=-11/3$), 
which conserves kinetic energy and runs to low $k$,
from the stirring scale up to the system scale; 
and a $-3$ cascade ($\alpha=-5$), 
which conserves mean-square vorticity and runs to high $k$,
from the stirring scale down to the dissipation scale. 
In a neutron star,
the stirring and system scales are approximately the same,
so the $-3$ cascade notionally covers a wider $k$ range than the $-5/3$ cascade. 
However, laboratory experiments indicate that the situation is more complicated.
For example, \citet{iid09} found 
$P(k) \propto k^{-4}$ for horizontal $k$ 
and $P(k) \propto k^{-5}$ for vertical $k$.
Vertical motions are suppressed, but the vertical cascade still plays 
a key role in routing energy through the system, by mediating the formation of
sheared stacks of pancake-like structures.
\citet{som86} showed experimentally that, even when
the largest scales are pancake-like, smaller scales remain isotropic and
Kolmogorov-like, perhaps due to the action of internal gravity waves
(which are weakly damped at the Prandtl numbers ${\rm Pr} \gg 1$
expected inside a neutron star). 
As a rule, stratified turbulence fossilizes into two dimensions when the activity
parameter $I = \varepsilon / (\nu N^2)$ drops below $\approx 7$
\citep{iid09},
where $N \approx 500 \, {\rm rad\,s^{-1}}$ is the Brunt-V\"{a}is\"{a}l\"{a} 
frequency in a typical neutron star in beta equilibrium \citep{van08}.
Some of the candidate astrophysical sources considered in \S\ref{sec:tur4}
satisfy this inequality, while others do not.
We confirm, with the help of order-of-magnitude estimates in \S\ref{sec:tur3c},
that the gravitational wave results
in \S\ref{sec:tur3} and \S\ref{sec:tur4} are probably insensitive to the
dimensionality of the turbulence,
at least for the values of $\alpha$ that one might reasonably expect in a neutron star.
Nonetheless,
we emphasize that the question is far from settled and needs to be studied
more carefully with more sophisticated, compressible numerical simulations.

\section{Gravitational wave signal
 \label{sec:tur3}}

\subsection{Current multipole moments
 \label{sec:tur3a}}
The gravitational wave strain measured by an observer
at a distance $d$ from a source can be written in the
transverse traceless gauge as a linear combination of
gravitoelectric (`mass') and gravitomagnetic (`current') multipoles;
see equation (4.3) of \citet{tho80}.
The latter components take the form
\begin{equation}
 h_{jk}^{\rm TT} 
 =
 \frac{G}{c^5 d} 
 \sum_{l=2}^\infty \sum_{m=-l}^l
 \frac{\partial^l S^{lm}(t)}{\partial t^l}
 T_{jk}^{{\rm B2},lm}~.
\label{eq:tur21}
\end{equation}
In equation (\ref{eq:tur21}), 
$S^{lm}(t)$ denotes the $(l,m)$-th current multipole moment,
written as a function of the retarded time $t$,
and $T_{jk}^{{\rm B2},lm}$ denotes the associated
gravitomagnetic tensor spherical harmonic,
which describes the angular dependence (or beam pattern)
of the radiation field.
For a Newtonian source (slow internal motions, weak internal gravity),
like a differentially rotating neutron star,
the current multipole moments assume the form
\begin{eqnarray}
 S^{lm}
 & = &
 - \frac{32\pi}{(2l+1)!!}
 \left[
  \frac{(l+2)(2l+1)}{2(l-1)(l+1)}
 \right]^{1/2}
 \nonumber \\
 & &
 \times
 \int d^3{\bf x} \, r^{l-1} 
 ({\bf x}\times \rho{\bf v})
 \cdot
 {\bf Y}^{l-1,lm\ast}
\label{eq:tur22}
 \\ 
 & = &
 - \frac{32\pi}{(2l+1)!!}
 \left[
  \frac{l+2}{2l(l-1)(l+1)}
 \right]^{1/2}
 \nonumber \\
 & &
 \times
 \int d^3{\bf x} \, r^l
 {\bf x} \cdot {\rm curl}(\rho{\bf v})
 Y^{lm\ast}
\label{eq:tur23}
\end{eqnarray}
where ${\bf Y}^{l-1,lm}$ is a vector spherical harmonic
of pure orbital type,
$Y^{lm}$ is a scalar spherical harmonic,
and $\rho({\bf x})$ is the fluid mass density.
Equation (\ref{eq:tur22}) corresponds exactly to equation (5.27b)
in \citet{tho80}.
Equation (\ref{eq:tur23}) is derived from (\ref{eq:tur22})
by expressing the vector harmonic in terms of gradients of scalar harmonics,
viz.\
\begin{equation}
 [l(2l+1)]^{1/2} {\bf Y}^{l-1,lm}
 =
 r\nabla Y^{lm} + l {\bf\hat x} Y^{lm}~,
\label{eq:tur24}
\end{equation}
and then integrating by parts
\citep{was09}.
Physically, therefore, the current multipole arises from the 
magnetic component of the velocity field 
[equation (\ref{eq:tur22})]
or, equivalently, from the radial component of the vorticity field
[equation (\ref{eq:tur23})].
We only consider incompressible turbulence in this paper
(see \S\ref{sec:tur2}), for which the mass multipoles vanish.

For the sake of simplicity, we take $\rho$ to be uniform,
i.e.\ $\rho=3M_\ast/(4\pi R_\ast^3)$,
where $M_\ast$ is the total mass of fluid in the stellar interior;
cf.\ \citet{was09}.
A more realistic assumption is that $\rho$ is incompressible
(subsonic flow) but stratified gravitationally.
However, we avoid stratification in this first pass at the problem
because we wish to exploit the scalings for 
isotropic Kolmogorov turbulence in \S\ref{sec:tur2},
which assume uniform $\rho$.
As noted in \S\ref{sec:tur2},
the stratified problem is much harder.

\subsection{Root-mean-square wave strain
 \label{sec:tur3b}}
The mean wave strain at the observer is zero for stationary,
isotropic turbulence, as discussed in \S\ref{sec:tur2}.
Therefore, to assess detectability, we compute the
autocorrelation function
\begin{equation}
 C(\tau) 
 =
 \left\langle 
  h_{jk}^{\rm TT}(t) h_{jk}^{\rm TT}(t')^\ast
 \right\rangle~,
\label{eq:tur25}
\end{equation}
which is positive definite for $t=t'$,
reduces to the RMS wave strain
$h_{\rm RMS} = \langle | h_{jk}^{\rm TT}(t) |^2 \rangle^{1/2}$
for $t=t'$,
and is a function of $t$ and $t'$ only through the combination
$\tau=t'-t$ for stationary turbulence.
For the sake of simplicity, we compute $C(\tau)$
for a single multipole $(l,m)$.
By doing so, we circumvent the following complication:
an observer at a particular position, doing a realistic detection experiment,
sees cross terms in $C(\tau)$ which mix multipoles together
(e.g.\ $S^{20}S^{21\ast}$).
The cross terms only vanish when averaged over all possible 
observer positions
[by the orthonormality of $T_{jk}^{JS,lm}$;
see equation (2.36) of \citet{tho80}].
In this sense, our detectability estimates are conservative; 
the signal from one multipole is obviously a lower bound on the total signal.

The wave strain autocorrelation function is the ensemble average of a product 
of time derivatives of $S^{lm}$.
In the special case of stationary turbulence,
where it is always possible to (notionally) fix $t$ ($t'$)
at some instant during the average and exchange 
$d/dt'$ ($d/dt$) with $d/d\tau$ ($-d/d\tau$),
it is possible to simplify the ensemble average using the
following formula from turbulence theory,
e.g.\ equation (12) of \citet{gog07}:
\begin{equation}
 \left\langle 
  \frac{\partial S^{lm}(t)}{\partial t}
  \frac{\partial S^{lm}(t')}{\partial t'}
 \right\rangle
 =
 -\frac{d^2}{d\tau^2}
 \left\langle 
  S^{lm}(t) S^{lm}(t')
 \right\rangle~.
\label{eq:tur26}
\end{equation}
Upon combining 
(\ref{eq:tur21}), (\ref{eq:tur23}), (\ref{eq:tur25}),
and (\ref{eq:tur26}),
and noting that $|T_{jk}^{{\rm B2},lm}|^2 \leq 1$,
we obtain the maximum autocorrelation 
an optimally situated observer can detect:
\begin{eqnarray}
 C(\tau)
 & = &
 (-1)^l
 \frac{G^2 \rho^2}{c^{10} r^2}
 \left[
  \frac{32\pi}{(2l+1)!!}
 \right]^2
 \frac{l+2}{2l(l-1)(l+1)}
 \frac{d^{2l}}{d\tau^{2l}}
 \int_0^{R_\ast} dr \, r^{l+3}
 \int_0^{R_\ast} dr' \, (r')^{l+3}
 \nonumber \\
 & &
 \times
 \int d^2{\bf\hat x} \int d^2{\bf\hat x}'
 \left\langle
  Y^{lm\ast}({\bf\hat x}) {\bf\hat x} \cdot {\bf \omega}({\bf x},t)
  Y^{lm}({\bf\hat x}') {\bf\hat x}' \cdot {\bf \omega}({\bf x}',t')^\ast
 \right\rangle~.
\label{eq:tur27}
\end{eqnarray}
Upon substituting (\ref{eq:tur3})--(\ref{eq:tur5}) into (\ref{eq:tur27}),
to deal with the ensemble average,
and performing the ${\bf k}'$ integral over the delta function,
we arrive at the expression
\begin{eqnarray}
 C(\tau)
 & = &
 (-1)^l
 \frac{G^2 \rho^2}{c^{10} r^2}
 \left[
  \frac{32\pi}{(2l+1)!!}
 \right]^2
 \frac{l+2}{2l(l-1)(l+1)}
 \frac{d^{2l}}{d\tau^{2l}}
 \int_0^{R_\ast} dr \, r^{l+3}
 \int_0^{R_\ast} dr' \, (r')^{l+3}
 \nonumber \\
 & &
 \times
 \int \frac{dk\,k^4}{(2\pi)^3}
  V P(k) \exp[-\pi \eta(k)^2\tau^2/4]
 \int d^2{\bf\hat x} \int d^2{\bf\hat x}' 
 \nonumber \\
 & &
 \times
  \int d^2{\bf\hat k} \,
  Y^{lm\ast}({\bf\hat x}) {\hat x}_i \exp(i{\bf k}\cdot{\bf x})
  Y^{lm}({\bf\hat x}') {\hat x}'_j \exp(-i{\bf k}\cdot{\bf x}')
 (\delta_{ij} - {\hat k}_i {\hat k}_j)~.
\label{eq:tur28a}
\end{eqnarray}

Equation (\ref{eq:tur28a}) can be simplified analytically
in a number of ways. Here, we elect to expand the two
plane-wave factors in terms of scalar spherical harmonics, viz.
\begin{equation}
 e^{i{\bf k}\cdot{\bf q}}
 =
 4\pi
 \sum_{L=0}^\infty i^L j_L(kq)
 \sum_{M=-L}^L Y^{LM\ast}({\bf\hat q}) Y^{LM}({\bf\hat k})~,
\label{eq:tur9}
\end{equation}
and exploit the orthogonality properties of the spherical
harmonics to make progress.
In (\ref{eq:tur9}),
$j_L$ denotes a spherical Bessel function of the first kind.
The six angular integrals
(over ${\bf \hat x}$, ${\bf \hat x}'$, and ${\bf \hat k}$)
now factorize easily,
and we arrive at the final expression for $C(\tau)$,
\begin{eqnarray}
 C(\tau)
 & = &
 (-1)^l
 \frac{G^2 \rho^2}{c^{10} r^2}
 \left[
  \frac{32\pi}{(2l+1)!!}
 \right]^2
 \frac{l+2}{\pi l(l-1)(l+1)}
 \frac{d^{2l}}{d\tau^{2l}}
 \nonumber \\
 & &
 \times
 \int_0^{R_\ast} dr \, r^{l+3}
 \int_0^{R_\ast} dr' \, (r')^{l+3}
 \int dk\, k^4 V P(k) \exp[-\pi \eta(k)^2 \tau^2 / 4]
 \nonumber \\
 & &
 \times
 \sum_{L=0}^\infty \sum_{L'=0}^\infty \sum_{M=-L}^L \sum_{M'=-L'}^{L'}
 i^L j_L(kr) (-i)^{L'} j_{L'}(kr')
 K_i^{lmLM} K_j^{lm L' M'\ast} N_{ij}^{LML' M'}~,
\label{eq:tur28b}
\end{eqnarray}
with
\begin{eqnarray}
 K_i^{lmLM}
 & = &
 \int d^2{\bf\hat x} \,
 Y^{lm\ast}({\bf\hat x}) {\hat x}_i Y^{LM\ast}({\bf\hat x})~,
 \\
\label{eq:tur29}
 N_{ij}^{LML' M'}
 & = &
 \int d^2{\bf\hat k} \,
 Y^{LM}({\bf\hat k}) Y^{L'M'\ast}({\bf\hat k})
 (\delta_{ij} - {\hat k}_i {\hat k}_j)~.
\label{eq:tur30}
\end{eqnarray}
One can evaluate $K_i$ and $N_{ij}$ by writing ${\hat x}_i$
and ${\hat k}_i$ in terms of $Y^{10}$ and $Y^{1,\pm1}$
and using Clebsch-Gordan coefficients to evaluate the
triple products. For example, $K_i$ is nonzero only if
$L=l\pm1$ and $|M+m|\leq 1$.
In what follows, however, we do the integrals directly
with the help of a symbolic algebra package.

\subsection{Quadrupole
 \label{sec:tur3c}}
Let us begin by specializing to the case $l=m=2$, 
where the sums in (\ref{eq:tur28b}) are nonzero
for $L=1,3$ and $L'=1,3$ only.
[The same is true for $(l,m)=(2,1)$.]
Doing the $r$ and $r'$ integrals and $\tau$ derivatives, 
we find
\begin{eqnarray}
 C(\tau)
 & = &
 \frac{256 \pi^3 G^2 \rho^2}{5625 c^{10} d^2}
 \int dk \, k^{-8} V P(k) \exp[-\pi \eta(k)^2 \tau^2 /4]
 \nonumber \\
 & &
 \times
 \eta(k)^4 
 [12 - 12\pi\eta(k)^2 \tau^2 + \pi^2 \eta(k)^4 \tau^4 ]
 [ \psi^{51}(kR_\ast) + \psi^{53}(kR_\ast) ]^2~.
\label{eq:tur31}
\end{eqnarray}
The integral defined by
\begin{equation}
 \psi^{ab}(x)
 =
 \int_0^x d\xi \xi^a j_b(\xi)~,
\label{eq:tur32}
\end{equation}
when combined with $P(k)$ through the $k$ integral
in (\ref{eq:tur31}),
expresses mathematically the fact that the wave strain is an
incoherent sum of randomly phased eddy motions,
a subset of which are wavenumber-matched to the multipole moment
under consideration.
It can be shown that $|C(\tau)|^{1/2}$ is roughly proportional 
to the square root of the mean number of eddies at the relevant scale
\citep{was09}.

The maximum RMS wave strain is obtained for $\tau=0$,
when there is no turbulent dephasing.
The $k$ integral in equation (\ref{eq:tur31}) runs from
the stirring to the dissipation scale,
i.e.\ from $k_{\rm s}$ in (\ref{eq:tur7a})
to  $k_{\rm d}$ in (\ref{eq:tur7b}).
The $\psi$ function is oscillatory, but its envelope grows 
$\propto k^3$ for $a=5$.
Hence,
if $\eta(k)$ and $P(k)$ are given by
(\ref{eq:tur6}) and (\ref{eq:tur7}) respectively,
the integrand scales $\propto k^{-3}$ overall
in the limit $\tau\rightarrow 0$.
{\em The RMS wave strain is therefore dominated by motions
at the stirring scale.}
Under these circumstances, with $k_{\rm s}\ll k_{\rm d}$,
the result is
\begin{equation}
 h_{\rm RMS}^2
 =
 \frac{0.59 G^2 \rho^2 R_\ast^8 \varepsilon^2}{c^{10} d^2}~.
\label{eq:tur33}
\end{equation}
If the turbulence is powered by differential rotation,
the specific (per unit mass) energy input per unit time is
$\varepsilon = R_\ast^2 (\Delta\Omega)^3$
\citep{lan59},
\footnote{
This choice of $\varepsilon$ is conservative. 
Other possible choices,
e.g.\ $\varepsilon = R_\ast^2 \Omega^2 \Delta\Omega$,
imply a higher gravitational wave strain.
}
from which we obtain
\begin{equation}
 h_{\rm RMS}
 =
 5\times 10^{-28}
 \left(
  \frac{M_\ast}{1.4M_{\odot}}
 \right)
 \left(
  \frac{R_\ast}{10\,{\rm km}}
 \right)^3
 \left(
  \frac{d}{1\,{\rm kpc}}
 \right)^{-1}
 \left(
  \frac{\Delta\Omega}{10\,{\rm rad\,s^{-1}}}
 \right)^3~.
\label{eq:tur34}
\end{equation}
The Rossby number is defined as ${\rm Ro}=\Delta\Omega/\Omega$,
as in \S\ref{sec:tur2}.

The above results confirm that we can reliably use the
quadrupole moment to estimate detectability.
As $h_{\rm RMS}$ is dominated by $k_{\rm s}$,
and the stirring scale is well matched to $l=2$,
we are not missing a lot of power emitted by eddies at
large $k$ and showing up at proportionally higher $l$.
On the debit side of the ledger,
the assumption of isotropy is least valid at the stirring scale,
as discussed in \S\ref{sec:tur2}.
Equation (\ref{eq:tur31}) makes it clear why this may
ultimately turn out to be a serious flaw:
$h_{\rm RMS}$ depends quite sensitively on the shape of $P(k)$.
For example, if the exponent of the power spectrum satisfies
$\alpha > -5/3$,
the dissipation scale governs $h_{\rm RMS}$,
not the stirring scale, 
and equation (\ref{eq:tur34}) would exhibit different scalings
with $\Omega$ and $\Delta\Omega$,
plus an explicit dependence on the
kinematic viscosity of the stellar interior.

Buoyancy arguably reduces the turbulence to two dimensions by suppressing
radial motion, as discussed in \S\ref{sec:tur2}.
However, order-of-magnitude estimates suggest that the gravitational wave signal
is insensitive to the dimensionality of the turbulence.
Evaluating equations (\ref{eq:tur28b}) and (\ref{eq:tur31}) for $\alpha=-5$,
appropriate for the vorticity-conserving $-3$ cascade postulated by \citet{kra67},
we find
that $h_{\rm RMS}$ changes by less than $20\%$ for $k_{\rm s} \ll k_{\rm d}$.
In reality, because $h_{\rm RMS}$ is dominated by motions near $k_{\rm s}$, 
where the forward and reverse cascades cross,
$\alpha$ effectively lies closer to $-5/3$ than to $-3$,
even in the two-dimensional model of \citet{kra67},
and the change is even smaller.
In all cases, the power-law dependences on $\varepsilon$ and the other quantities
in equation (\ref{eq:tur33}) are universal, as long as we have $\alpha < -5/3$;
the shape of the turbulent spectrum affects just the numerical prefactor 
in equation (\ref{eq:tur33}), and even then only weakly,
unless we have $\alpha > -5/3$, whereupon the dissipation scale dominates the signal. 
Theory and experiment are united in deeming spectra shallower than $P(k) \propto k^{-11/3}$
to be extremely rare in nature;
certainly, buoyancy acts in the contrary direction to steepen $P(k)$.
As noted in \S\ref{sec:tur2}, although neutron stars are strongly stratified,
the activity parameter 
$I = \varepsilon/(\nu N^2) 
 \approx 40 (\Delta\Omega / 1\,{\rm rad\,s^{-1}})^3 (\nu/10\,{\rm m^2\,s^{-1}})^{-1}
 (N/500 \, {\rm rad\,s^{-1}})^{-2}$
falls below the fossilization threshold $I\approx 7$ 
\citep{iid09}
for 
$\Delta\Omega < 0.6
 (\nu/10\,{\rm m^2\,s^{-1}})^{1/3}
 (N/500 {\rm rad\,s^{-1}})^{2/3}$
\, ${\rm rad\,s^{-1}}$.
Some of the candidate sources discussed below in \S\ref{sec:tur4} 
do not satisfy this inequality (see Table \ref{tab:tur1});
that is, hydrodynamic turbulence in these sources may be effectively
three-dimensional despite strong stratification,
lending extra support to the results in this section.

\subsection{Higher multipoles
 \label{sec:tur3d}}
As a general rule, the current multipole moment depends
increasingly strongly on $k_{\rm d}$ as $l$ increases,
because higher multipoles match better to small-scale eddies.
This effect is communicated through the $\psi$ function 
combined with $P(k)$,
e.g.\ as in equation (\ref{eq:tur31}).
Whenever $l$ increases by one,
the $\tau$ derivatives in (\ref{eq:tur28b}) bring down
two extra powers of $\eta(k)$,
the $r$ and $r'$ integrals contribute an extra factor of $k^2$,
and the $\psi$ functions contribute new $k$ factors too.
For $l=3$, we find
\begin{eqnarray}
 h_{\rm RMS}^2
 & = &
 \frac{8 \pi^3 G^2 \rho^2 \varepsilon^{8/3}}{7203 c^{12} d^2}
 \int_{k_{\rm s}}^{k_{\rm d}} dk \, k^{-29/3}
 [ \psi^{62}(kR_\ast) + \psi^{64}(kR_\ast) ]^2
\label{eq:tur35}
 \\
 & = &
 \frac{0.41 G^2 \rho^2 R_\ast^{26/3} \varepsilon^{8/3}}
  {c^{12} d^2}~,
\label{eq:tur36}
\end{eqnarray}
with $k_{\rm s} \ll k_{\rm d}$.
The integrand in (\ref{eq:tur35}) scales $\propto k^{-5/3}$,
so $h_{\rm RMS}$ for $l=3$ is dominated by motions
at the stirring scale, just like for $l=2$.
The ratio of the $l=3$ and $l=2$ wave strains is
$0.83 (R_\ast\Omega/c) {\rm Ro}$,
implying that the $l=3$ radiation is significantly weaker
for any realistic neutron star rotating slower than
centrifugal breakup.

For $l\geq 4$, one finds that $h_{\rm RMS}$ is dominated
by motions at the dissipation scale, not the stirring scale.
Consequently, the dependence of $h_{\rm RMS}$ on
$\Omega$ and $\Delta\Omega$ also changes,
and a new, explicit dependence on the kinematic viscosity appears.
For example, for $l=4$, one obtains
$h_{\rm RMS} \propto 
 \rho R_\ast^5 \varepsilon^{5/3} k_{\rm d}^{1/3} d^{-1}
 \propto \nu^{-1/4}$,
and the ratio of the $l=4$ and $l=2$ wave strains is
$\approx (R_\ast \Omega/c)^2 {\rm Ro}^{2} (k_{\rm d}R_\ast)^{1/3}$.
In view of the Kolmogorov scaling 
$k_{\rm d}/k_{\rm s} = {\rm Re}^{3/4}$
[see equation (17) of \citet{kos02}],
we conclude that the $l=4$ radiation
is weaker than the $l=2$ radiation for
${\rm Re} \lesssim (R_\ast \Omega/c)^{-8} {\rm Ro}^{-8}$,
which is always satisfied except possibly in a strongly sheared
millisecond pulsar, e.g. one born recently in a supernova.

\subsection{Decoherence time
 \label{sec:tur3e}}
The decoherence time $\tau_{\rm c}$ is the time
that must elapse before the instantaneous wave strains
$h_{jk}^{\rm TT}(t)$ and $h_{jk}^{\rm TT}(t')$ 
become statistically uncorrelated.
We define it to be the value of $\tau=t'-t$ at which $C(\tau)$,
defined as the ensemble-averaged correlator in (\ref{eq:tur25}), 
decreases to some fixed fraction (say, one quarter)
of its maximum (at $\tau=0$).

The results in \S\ref{sec:tur3c} and \S\ref{sec:tur3d}
demonstrate that $h_{\rm RMS}$ is generated predominantly
by the stress-energy in motions at the stirring scale.
It is tempting, therefore, to predict that the signal decoheres
on the eddy turnover time at the stirring scale,
i.e.\ the rotation period of the star multiplied by ${\rm Ro}$
\citep{kos02}.
In fact, the situation is potentially more complicated.
Looking at equation (\ref{eq:tur31}), 
which describes how $C(\tau)$ decreases with $\tau$,
we see that the integrand contains three terms,
which scale 
$\propto k^{-3}$, $k^{-5/3}$, and $k^{-1/3}$.
The first, positive term is the only one which contributes to
the maximum of $h_{\rm RMS}$,
i.e.\ equation (\ref{eq:tur33});
its integral is dominated by $k_{\rm s}$, as discussed above.
The second, negative term is the only one which can reduce $C(\tau)$ and
cause decoherence.
Its integral is also dominated by $k_{\rm s}$.
The integral of the third, positive term involves both $k_{\rm s}$ and $k_{\rm d}$
but is dominated by the former (see next paragraph).
Hence the final result for $\tau_{\rm c}$ contains information
about both $k_{\rm s}$ and $k_{\rm d}$ in general.
But, for the $l=2$ and $l=3$ signals, it turns out that the dependence on $k_{\rm d}$ 
is very weak, unlike for higher $l$.
This result is crucial for the question of detectability,
as explained in \S\ref{sec:tur4}.

\begin{figure}
\plotone{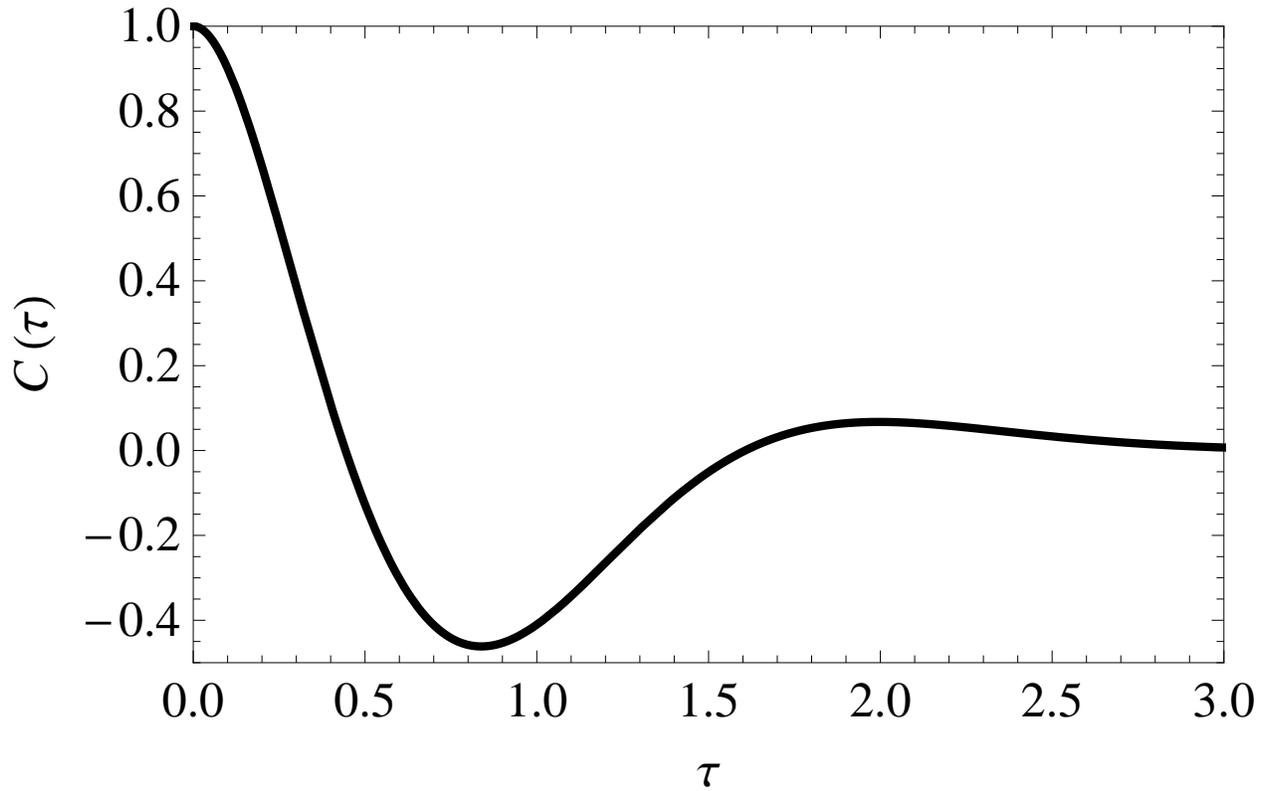}
\caption{
Wave strain autocorrelation function $C(\tau)$, defined by (\ref{eq:tur37}),
normalized by the mean-square wave strain $h_{\rm RMS}^2$,
as a function of the time lag $\tau$, 
normalized by the maximum eddy turnover time $\eta(k_{\rm s})^{-1}$
at the stirring scale, for $k_{\rm d} /k_{\rm s} = 10^3$.
}
\label{fig:tur4}
\end{figure}

By graphing $C(\tau)$ numerically in Figure \ref{fig:tur4}, 
we find that it falls to one quarter of its maximum over a time comparable to
the eddy turnover time at the stirring scale, $\eta(k_{\rm s})^{-1}$.
This graphical task is tricky, because the square of the $\psi^{ab}$ functions
oscillates rapidly for $k_{\rm d} \gg k_{\rm s}$.
However, we can obtain an excellent analytic approximation by averaging over
many cycles of the fast oscillation. 
Upon writing $[\psi^{51}(x) + \psi^{53}(x)]^2 \approx 25 x^6/2$
plus fast oscillations $\propto \cos(2x)$ in the regime $x\gg 1$,
we arrive at the formula
\begin{eqnarray}
 \frac{C(\tau)}{h_{\rm RMS}^2} 
 & = & 
 \left[
  1 - \frac{ 7\pi \eta(k_{\rm s})^2 \tau^2 }{2}
 \right]
  \exp \left[ - \frac{\pi \eta(k_{\rm s})^2 \tau^2}{4} \right]
 \nonumber \\
 & & 
 + 2 \pi^2 \eta(k_{\rm s})^3 \tau^3
 \left\{
  {\rm Erf} \left[ \frac{\pi^{1/2}\eta(k_{\rm s})\tau}{2} 
    \left( \frac{k_{\rm d}}{k_{\rm s}} \right)^{2/3} \right]
 -
  {\rm Erf} \left[ \frac{\pi^{1/2}\eta(k_{\rm s})\tau}{2} \right]
 \right\}~,
\label{eq:tur37}
\end{eqnarray}
where ${\rm Erf}(x)$ symbolizes the error function.
Equation (\ref{eq:tur37}) is accurate to $\approx 12\%$ across the full range of $\tau$
for the parameter range under consideration.
The dependence on $k_{\rm d}/k_{\rm s} \gg 1$ in the third term is weak,
although this stops being true for the rare case of non-Kolmogorov spectra with
$\alpha > -5/3$, where $k_{\rm d}$ dominates
(see \S\ref{sec:tur3c}).
From (\ref{eq:tur6}), (\ref{eq:tur7a}), (\ref{eq:tur7b}), and (\ref{eq:tur37}),
with
$\varepsilon=R_\ast^2 (\Delta\Omega)^3$,
we arrive at the following expression for the decoherence time
corresponding to the half-strain point:
\footnote{
It should be noted that $\eta(k)$ does not always follow
(\ref{eq:tur7a}) near the stirring scale
\citep{gog07}, as discussed in \S\ref{sec:tur2}.
We aim to refine (\ref{eq:tur40}) in future work.
}
\begin{eqnarray}
 \tau_{\rm c}
 & = &
 0.35 \eta(k_{\rm s})^{-1}
\label{eq:tur40}
 \\
 & = &
 26\,{\rm ms}
 \left(
  \frac{\Delta\Omega}{10\,{\rm rad\,s^{-1}}}
 \right)^{-1}~.
\label{eq:tur41}
\end{eqnarray}

The shear viscosity in a neutron star is given by
the neutron-neutron scattering formula derived by \citet{cut87},
viz.\
$\nu = 10 \,
 (\rho/6\times 10^{17}\,{\rm kg\,m^{-3}})^{5/4}
 (T/10^8\,{\rm K})^{-2} \, {\rm m^2\,s^{-1}}$,
where $T$ is the temperature of the stellar interior.
From equation (\ref{eq:tur7b}), we get
$k_{\rm d} \gg k_{\rm s}$ for fiducial values of $\rho$ and $T$.
In a protoneutron star, discussed in \S\ref{sec:tur4},
the bulk viscosity ($\propto T^8$) exceeds the shear viscosity
\citep{saw89}.
Under these conditions, small compressions enhance the turbulent dissipation rate,
effectively reducing $k_{\rm d}$.
Numerical simulations of a compressible HVBK superfluid
(outside the scope of this paper)
are needed to quantify this effect properly.
However, the foregoing gravitational wave calculations continue to hold
to a good approximation,
provided that the stirring and dissipation scales remain moderately well separated
[e.g.\ $k_{\rm d}/k_{\rm s} \gtrsim 5$ in equation (\ref{eq:tur37})].

\section{Detectability and astrophysical implications
 \label{sec:tur4}}
In this paper, we calculate analytically the current multipole
moments generated by vorticity fluctuations in high-${\rm Re}$
turbulence in a differentially rotating neutron star.
We derive an analytic expression [equation (\ref{eq:tur31})]
for the wave strain autocorrelation function $C(\tau)$ of the
resulting, stochastic gravitational wave signal
to leading (quadrupole) order,
in terms of the turbulent power spectrum
and eddy turnover time spectrum,
and show that $C(\tau)$ is governed by motions at the
stirring scale.
From the same equation, we also compute 
the root-mean-square wave strain $h_{\rm RMS}$ and
the decoherence time
$\tau_{\rm c}$ of the signal
and show
that $\tau_{\rm c}$ approximately equals the
eddy turnover time at the stirring scale
(usually much longer than the rotation period).
Convenient formulas for $h_{\rm RMS}$ and $\tau_{\rm c}$,
scaled in terms of astrophysical parameters,
are presented in (\ref{eq:tur34}) and (\ref{eq:tur41}) respectively.

We can use these formulas to assess qualitatively whether
the stochastic signal can be detected by existing and
planned long-baseline gravitational wave interferometers.
Clearly, neutron star turbulence imposes a fundamental,
unavoidable, astrophysical noise floor on continuous-wave searches 
for neutron stars,
e.g.\ \citet{abb09}.
But how seriously does it pollute such searches?
To answer this question, we note two things.
First, most neutron stars are either rotating too slowly
or with too little shear to cause trouble,
according to (\ref{eq:tur34}),
at least at the sensitivities anticipated for Enhanced and Advanced LIGO.
Second, for the small subset of neutron stars that are
potentially powerful stochastic emitters,
the signal decoherence time is relatively short;
indeed, for $\Delta\Omega \sim \Omega$ and the fastest rotators,
$\tau_{\rm c}$ approaches (without ever quite reaching) 
the sampling time of LIGO-like interferometers .
Equations (\ref{eq:tur34}) and (\ref{eq:tur41}) imply
that $\tau_{\rm c}$ decreases as $h_{\rm RMS}$ increases.

Table \ref{tab:tur1} illustrates these conclusions by listing
$h_{\rm RMS}$ and $\tau_{\rm c}$ for several realistic categories
of neutron star sources.
\begin{enumerate}
\item
{\em Protoneutron stars.}
The stellar angular velocity profile
is a key output of radiation (magneto)hydrodynamics simulations 
of core-collapse supernovae
\citep{ott06,bur07}.
All progenitor models tested so far
lead to strong differential rotation.
Looking for example at the fastest rotators in 
Figures 8, 9, and 15 of \citet{ott06}
or Figure 14 of \citet{bur07},
captured $0.2\,{\rm s}$ after core bounce
in a two-dimensional, unmagnetized explosion,
we see that $\Omega$ decreases gradually from
$\approx 3\times 10^3\,{\rm rad\,s^{-1}}$ at $r=3\,{\rm km}$
to
$\approx 1\times 10^3\,{\rm rad\,s^{-1}}$ at $r=30\,{\rm km}$
(enclosed mass $\approx 1.2 M_{\odot}$),
before dropping steeply to
$\approx 20\,{\rm rad\,s^{-1}}$ at $r=300\,{\rm km}$
(enclosed mass $\approx 1.7 M_{\odot}$).
\footnote{
Strongly magnetized models rotate $\sim 10$ times slower
\citep{heg05,bur07};
cf.\ millisecond protomagnetar engine for 
long-duration gamma-ray bursts
\citep{buc07}.
} 
Conservatively,
for a hypothetical supernova in the Milky Way,
the above figures imply
$h_{\rm RMS} = 9\times 10^{-21}$
and $\tau_{\rm c}=0.1\,{\rm ms}$
(first line of Table \ref{tab:tur1}).
LIGO is capable of detecting such stochastic emission,
because $\tau_{\rm c}$ is greater than the sampling time
of the interferometer ($\approx 60\,\mu{\rm s}$),
but the signal partially decoheres.
The prospects are brighter if the protoneutron star
rotates slower and with less shear,
and the gravitational radiation emanates mainly from 
the extended envelope 
(second line of Table \ref{tab:tur1}),
so that lower $h_{\rm RMS}$ ($5\times 10^{-22}$)
is traded for higher $\tau_{\rm c}$ ($3\,{\rm ms}$).
Detection by Advanced LIGO is then possible in principle,
e.g.\ via a cross-correlation search
\citep{dhu08}.
The signal persists for $\gtrsim 10^2\,{\rm s} \gg \tau_{\rm c}$
before the differential rotation dissipates
and/or the protoneutron star spins down,
e.g.\ due to $r$-modes, fallback, or a magnetized wind
\citep{lai01,ott06}.
\item
{\em Glitchers.}
Discontinuous spin-up events (`glitches') observed
in some rotation-powered pulsars
are adduced as evidence that neutron stars
rotate differentially;
the nuclear lattice spins down electromagnetically,
lagging the superfluid due to vortex pinning
\citep{and75}.
The fractional jump in angular velocity ranges from
$10^{-11}$ to $10^{-4}$ across the pulsar population.
The surprising absence of a `reservoir effect'
(i.e.\ glitch size $\propto$ time elapsed
since the preceding glitch)
in most objects
implies that the observed spin ups are a small percentage
of the underlying shear
\citep{mel08,war08}.
It is therefore safe to expect ${\rm Ro} \geq 10^{-4}$ 
in some pulsars,
although the proportion is hard to quantify.
Two examples are given in Table \ref{tab:tur1}:
an adolescent, Vela-like pulsar (age $\sim 10^4\,{\rm yr}$),
which spins relatively slowly ($\sim 10\,{\rm Hz}$)
but undergoes relatively large glitches
(fractional jump $\sim 10^{-6}$);
and a young, Crab-like pulsar (age $\sim 10^3\,{\rm yr}$),
which spins faster ($\sim 30\,{\rm Hz}$)
but undergoes smaller gliches
(fractional jump $\sim 10^{-8}$).
Assuming that the typical glitch resets $\sim 0.01\%$ of the underlying shear,
we find that the decoherence time 
(0.3$\,{\rm s}$ to 10$\,{\rm s}$) matches well to
the LIGO pass band,
but the signal is too weak ($h_{\rm RMS} \lesssim 10^{-30}$)
to be detected by interferometers under development,
or to pollute continuous-wave searches targeted at
glitching pulsars
\citep{van08}.
\footnote{
The persistent, stochastic, gravitational wave emission 
from shear-driven hydrodynamic turbulence is unrelated to the
putative burst emission from the glitches themselves.
}
\item
{\em Accretors.}
The hydromagnetic accretion torque acting on a compact star
in a mass-transfer binary is often comparable to, or greater than, 
the electromagnetic spin-down torque acting on an isolated pulsar
\citep{har08}
and fluctuates by several per cent daily,
causing X-ray variability.
Accreting objects are therefore likely to rotate
differentially, with Rossby numbers comparable to,
or greater than, those of isolated glitching pulsars.
Two examples are given in Table \ref{tab:tur1}:
a standard, accreting millisecond pulsar (fifth line),
and an accreting white dwarf with an ONeMg core (sixth line)
in a system on the verge of accretion-induced collapse
\citep{bla98,des07,met08}.
The angular velocity profile of the white dwarf before
(and after) collapse is plotted in Figure 12 of \citet{des06}.
In both cases, $h_{\rm RMS} \lesssim 10^{-26}$ falls below
the threshold for detection by Advanced LIGO but may approach
the sensitivity of the next generation of interferometers.
\item
{\em Nearby neutron stars.}
The nearest radio pulsar discovered to date is 
PSR J0108$-$1431, with $d=85\,{\rm pc}$
\citep{tau94}.
The nearest millisecond pulsar is PSR J0437$-$4715,
with $d=157\,{\rm pc}$
\citep{ver08}.
However, 
many radio-quiet neutron stars with $d<85\,{\rm pc}$
should reside in the Solar neighborhood;
the evidence is both observational
[e.g.\ radio-quiet X-ray point sources like the `Magnificent Seven',
some with parallaxes;
see \citet{pop03} and references therein]
and theoretical
[e.g.\ population synthesis models predict $\sim 10^9$
compact objects in the Milky Way;
see \citet{kie08} and references therein].
If the nearest objects have $d\sim 10\,{\rm pc}$ and
rotate reasonably fast, they represent promising LIGO candidates.
\footnote{
Our inability to measure an ephemeris from radio timing
observations does not affect cross-correlation searches for the
stochastic radiation from neutron star turbulence,
although it is a major obstacle to coherent continuous-wave searches.
}
For example,
Table \ref{tab:tur1} quotes $h_{\rm RMS}$ for two nearby
isolated pulsars with ${\rm Ro}=10^{-1}$.
The millisecond pulsar in particular is a bright source,
although it suffers from the usual drawback of fast rotators:
$\tau_{\rm c}$ is short.
Although it is sometimes assumed that millisecond pulsars do not
rotate differentially, because they are not seen to glitch,
it is equally possible that they glitch strongly but infrequently,
because the electromagnetic spin-down torque is weak
\citep{mel08}.
\end{enumerate}

\begin{table}
\begin{tabular}{lrrrrrr} \hline
 Type & $d$ (kpc) & $\Omega$ (${\rm rad\,s^{-1}}$) & ${\rm Ro}$ & 
  $h_{\rm RMS}$ & $\tau_{\rm c}$ (s) &
  $|\dot{\Omega}_{\rm GW}|$ (${\rm rad\,s^{-2}}$) 
\\ \hline
 Protoneutron star & & & & & & 
\\
 ($a$) core ($30\,{\rm km}$) &
  10 & $2\times10^3$ & $1$ & 
  $9\times 10^{-21}$ & $1\times 10^{-4}$ & 
  $3\times 10^{0}$
\\
  ($b$) envelope ($300\,{\rm km}$) &
  10 & $1\times10^3$ & $0.1$ & 
  $5\times 10^{-22}$ & $3\times 10^{-3}$ & 
  $1\times 10^{-6}$
\\ \hline
 Glitching pulsar & & & & & &
\\
  ($a$) Vela-like &
  1 & $1\times10^2$ & $10^{-2}$ & 
  $5\times 10^{-31}$ & $3\times 10^{-1}$ & 
  $3\times 10^{-27}$
\\
  ($b$) Crab-like &
  1 & $2\times10^2$ & $10^{-4}$ & 
  $4\times 10^{-36}$ & $1\times 10^{1}$ & 
  $4\times 10^{-41}$
\\ \hline
 Accretor & & & & & &
\\
  ($a$) millisecond pulsar &
  1 & $3\times10^3$ & $10^{-2}$ & 
  $1\times 10^{-26}$ & $9\times 10^{-3}$ & 
  $7\times 10^{-17}$
\\
  ($b$) white dwarf &
  0.1 & $0.2$ & $0.1$ & 
  $2\times 10^{-28}$ & $1\times 10^1$ & 
  $3\times 10^{-29}$
\\ \hline
 Nearby pulsar & & & & & & 
\\
 ($a$) fast rotator & 
  0.01 & $3\times10^3$ & $0.1$ & 
  $1\times 10^{-21}$ & $9\times 10^{-4}$ & 
  $7\times 10^{-9}$
\\
  ($b$) slow rotator & 
  0.01 & $2\times10^2$ & $0.1$ & 
  $4\times 10^{-25}$ & $1\times 10^{-2}$ & 
  $4\times 10^{-17}$
\\ 
\hline
\end{tabular}
\caption{
Candidate source categories.
}
\label{tab:tur1}
\end{table}

Even before detecting gravitational waves, we can use
existing radio timing observations of neutron stars
to test and constrain the model in this paper.
The radiation emitted by hydrodynamic turbulence
exerts a reaction torque,
which fluctuates in sign instantaneously but drains
rotational kinetic energy from the fluid over time.
The angular velocity of the star decreases
at the average rate
\citep{tho80,was09}
\begin{eqnarray}
 | \dot{\Omega}_{\rm GW} |
 & = &
 \frac{5G}{64\pi c^3 M_\ast R_\ast^2 \Omega}
 \sum_{l=2}^\infty \sum_{l=-m}^m 
 \left\langle
  \frac{1}{c^{2l}}
  \left|
   \frac{\partial^{l+1} S^{lm}}{\partial t^{l+1}}
  \right|^2
 \right\rangle
\label{eq:tur42}
\\
 & = &
 \frac{0.70 G \rho^2 R_\ast^{14/3} \varepsilon^{8/3}}
  {M_\ast c^7 \Omega}
\label{eq:tur42b}
\\
 & = &
 3\times 10^{-20}
 \left(
  \frac{M_\ast}{1.4 M_{\odot}}
 \right)
 \left(
  \frac{R_\ast}{10\,{\rm km}}
 \right)^{4}
 \left(
  \frac{\Omega}{10^3\,{\rm rad\,s^{-1}}}
 \right)^{-1}
 \left(
  \frac{\Delta\Omega}{10\,{\rm rad\,s^{-1}}}
 \right)^{8}
 {\rm rad\,s^{-2}}~.
\label{eq:tur43}
\end{eqnarray}
Equation (\ref{eq:tur43}) follows from (\ref{eq:tur42})
via the sequence of steps in \S\ref{sec:tur3a}--\S\ref{sec:tur3c}.
In the last column of Table \ref{tab:tur1},
we compute $|\dot{\Omega}_{\rm GW}|$ for the 
source categories discussed above.
In several instances, 
the fiducial value of $|\dot{\Omega}_{\rm GW}|$
is close to, or even greater than, the value of $|\dot{\Omega}|$ 
measured in radio timing experiments.
Already, this places constraints on the source parameters,
as the inequality $|\dot{\Omega}_{\rm GW}| \leq |\dot{\Omega}|$
must always hold for any specific object.
For example, the oldest radio millisecond pulsars are measured
to have $|\dot{\Omega}| \sim 10^{-15}\,{\rm rad\,s^{-2}}$
[see, for example, Table 10.1 in \citet{lyn98}].
Accreting millisecond pulsars in low-mass X-ray binaries have 
$|\dot{\Omega}| \lesssim 10^{-13}\,{\rm rad\,s^{-2}}$
during X-ray outbursts and
$|\dot{\Omega}| \sim 10^{-16}\,{\rm rad\,s^{-2}}$
between X-ray outbursts
\citep{har08}.
If these data are representative of the entire
millisecond pulsar population,
then we can start to rule out the existence of large shears 
like those quoted in the fifth and seventh entries of Table \ref{tab:tur1}.
Indeed,
equation (\ref{eq:tur43}) and the data combine to yield a 
{\em direct, observational upper limit on the rotational shear 
in any neutron star},
viz.\ 
\begin{equation}
 {\rm Ro} \leq
 4\times 10^{-2} 
 \left( 
  \frac{\Omega}{10^3\,{\rm rad\,s^{-1}}}
 \right)^{-7/8}
 \left( 
  \frac{|\dot{\Omega}|}{10^{-15}\,{\rm rad\,s^{-2}}}
 \right)^{1/8}~.
\label{eq:tur44}
\end{equation}

\begin{figure}
\plottwo{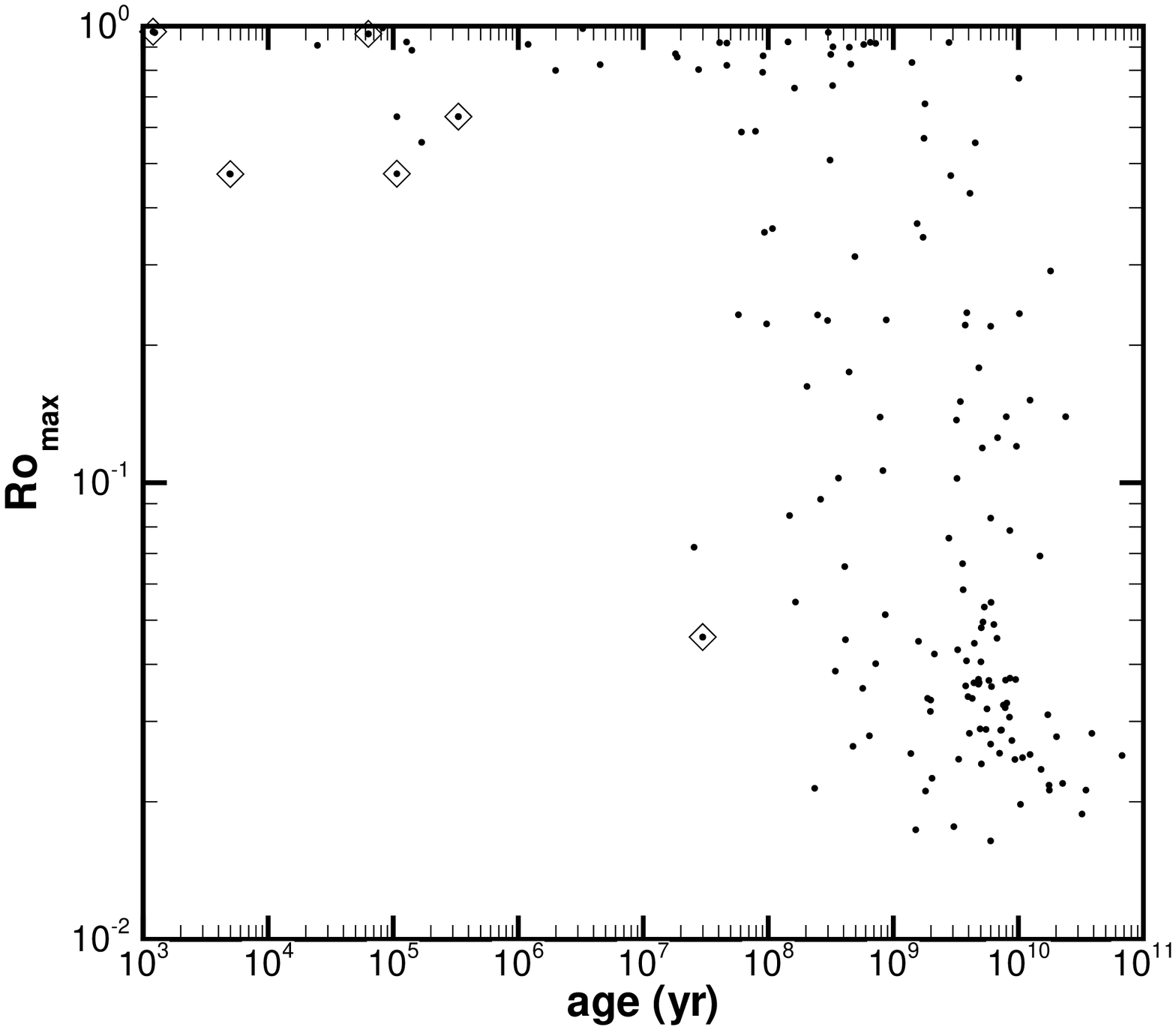}{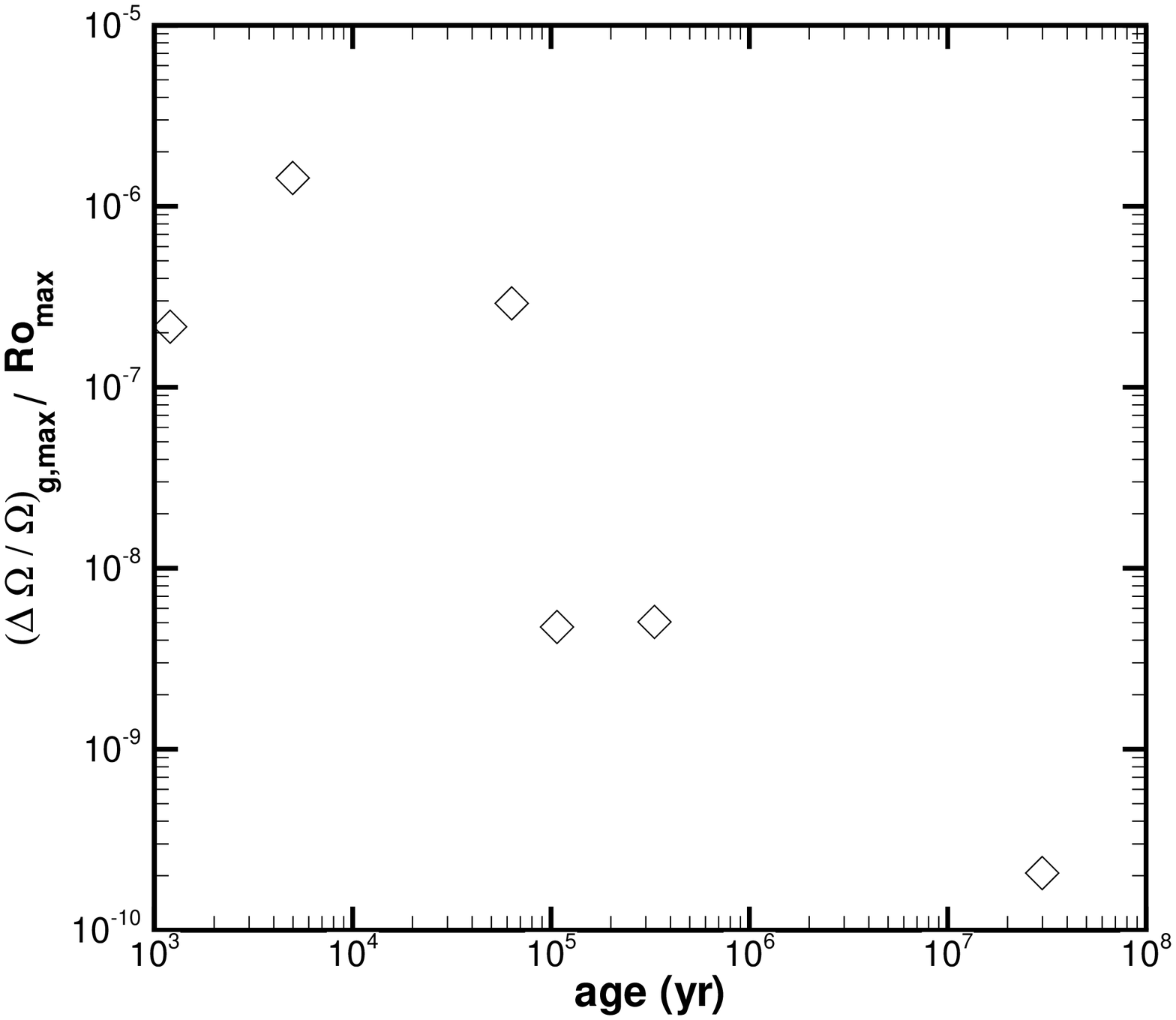}
\caption{
Upper limit on the angular shear in rotation-powered pulsars,
assuming gravitational wave spin down due to hydrodynamic turbulence.
{\em Left panel.}
Maximum Rossby number, ${\rm Ro}_{\rm max}$,
given by the right-hand side of equation (\ref{eq:tur44}),
versus characteristic age,
$\Omega/(2| \dot{\Omega} |)$ (in yr),
for objects with ${\rm Ro}_{\rm max} \geq 1$ 
in the ATNF Pulsar Catalog.
The points marked with diamonds indicate pulsars with a history
of glitch activity.
{\em Right panel.}
Maximum observed glitch size,
$(\Delta\Omega/\Omega)_{\rm g,max}$,
expressed as a fraction of ${\rm Ro}_{\rm max}$,
plotted versus characteristic age,
for the points marked with diamonds in the left panel.
The ratio should not exceed unity for any pulsar.
}
\label{fig:tur5}
\end{figure}

Equation (\ref{eq:tur44}) is an important result.
It allows the theory in this paper to be falsified,
if a glitching pulsar is discovered whose fractional
angular velocity jumps exceed
the right-hand side of (\ref{eq:tur44}).
We will investigate this application more thoroughly
in a forthcoming article.
Figure \ref{fig:tur5} gives a taste of what is possible.
In the {\em left panel} of Figure \ref{fig:tur5},
we plot ${\rm Ro}_{\rm max}$
[i.e.\ the right-hand side of (\ref{eq:tur44})]
versus spin-down age 
$\Omega/(2|\dot{\Omega}|)$
for all objects with ${\rm Ro}_{\rm max} \leq 1$ in the
Australia Telescope National Facility (ATNF)
Pulsar Catalog
\citep{man05}.
\footnote{
Objects with ${\rm Ro}_{\rm max} > 1$ do not place
a useful limit on the shear.
}
Clearly, millisecond pulsars with ages $\gtrsim 10^8\,{\rm yr}$
place strict limits on ${\rm Ro}$,
with ${\rm Ro}_{\rm max} \sim 10^{-2}$
in some cases.
Furthermore,
the pulsars marked with diamonds
have been observed to glitch at least once.
If the theory in this paper is correct,
then the largest angular velocity jump observed,
$(\Delta\Omega/\Omega)_{\rm g,max}$,
cannot exceed the underlying shear, $\Delta\Omega/\Omega$,
in each of the glitching pulsars;
in fact, it is expected to be much smaller,
given the absence of a reservoir effect.
As a test, we plot 
$(\Delta\Omega/\Omega)_{\rm g,max} / {\rm Ro}_{\rm max}$
versus spin-down age
in the {\em right panel} of Figure \ref{fig:tur5}.
In all cases, the plotted ratio is much smaller
than unity, consistent with theory.
\footnote{
The limits derived from Figure \ref{fig:tur5} are conservative.
In the standard vortex unpinning paradigm,
the fractional angular velocity jump observed in a large glitch 
can be related to the internal shear according to
$\sim (I_{\rm s}/I_{\rm c})(\Delta r/R) (\Delta\Omega/\Omega)$,
where $I_{\rm s}/I_{\rm c} \sim 10^2$ is the ratio of 
the superfluid and crustal moments of inertia, 
and $\Delta r / R \sim 10^{-6}$ is the normalized radial distance 
moved by the unpinned vortices
\citep{alp86}.
Arguably, therefore, the theory is challenged if
$(\Delta\Omega/\Omega)_{\rm g,max} / {\rm Ro}_{\rm max}$
exceeds $\sim 10^{-4}$ rather than unity,
a stronger test.
}
Future observational campaigns to simultaneously monitor 
large groups of pulsars, e.g.\ with multibeam
radio telescopes like the Square Kilometer Array,
will challenge the theory more keenly.

In summary, hydrodynamic turbulence imposes
a fundamental noise floor on gravitational wave observations
of neutron stars.
Its polluting effect is muted by partial decoherence in the
hectohertz band,
where current continuous-wave searches are concentrated,
but only for the fastest rotators with the strongest shear.
In addition, the mechanism sets a fundamental lower limit
on the spin-down rate $|\dot{\Omega}|$
and hence an observational upper limit on the
Rossby number ${\rm Ro}$, when combined with pulsar timing data.
These conclusions hold subject to one major caveat,
which is discussed at length in \S\ref{sec:tur2};
to wit, that $h_{\rm RMS}$ and $\tau_{\rm c}$ depend somewhat
on the exact shape of the turbulent power spectrum, $P(k)$.
It is known from laboratory experiments that $P(k)$ is modified
away from its isotropic, Kolmogorov form by the presence
of anisotropic global structures like turbulent boundary layers,
which are profoundly difficult to model in terrestrial contexts,
let alone in a neutron star.
Likewise, buoyancy modifies $P(k)$ by creating unequal
turbulent cascades parallel and perpendicular to the stratification direction,
in a fashion that is still not understood completely even in
controlled laboratory experiments.
The order-of-magnitude estimates in \S\ref{sec:tur3c} provide some comfort
that the gravitational wave results are insensitive to the above considerations,
but we emphasize that much work (e.g.\ compressible HVBK simulations)
still needs to be done to clarify the issue.

More complete detectability estimates referring to specific
search pipelines lie outside the scope of this paper.
Likewise, we defer calculating the detailed frequency spectrum 
of the stochastic signal, a substantial task.


\acknowledgments
The authors thank Ira Wasserman for privileged early access to an 
illuminating preprint \citep{was09},
Sterl Phinney for directing us to the work of Kosowsky and collaborators,
Scott Hughes for a valuable discussion on multipole expansions,
and the anonymous referee for spotting an error in our original calculation
of the decoherence time and for generally improving the manuscript.
We also acknowledge the substantial computing resources and support
provided by the Victorian Partnership for Advanced Computing,
which made the simulations in this paper possible.
AM thanks the Astronomy Department at Cornell and the 
LIGO Data Analysis Group at the California Institute of Technology
for their hospitality during the period when this work began,
and the LIGO Visitor Program for financial support.
CP acknowledges the support of the Max-Planck Society
(Albert-Einstein Institut).

\bibliographystyle{apj}
\bibliography{mpw06a}

\clearpage



\end{document}